\documentclass[preprint,showpacs,preprintnumbers,amsmath,amssymb]{revtex4}

\usepackage{graphicx}
\usepackage{dcolumn}
\usepackage{bm}

\usepackage{feynmf}
\usepackage{slashed}

\usepackage[utf8]{inputenc}

\begin{document}

\title{Vector-like Sneutrino Dark Matter}

\author{Yi-Lei Tang}
\thanks{tangyilei15@pku.edu.cn}
\affiliation{Center for High Energy Physics, Peking University, Beijing 100871, China}


\author{Shou-hua Zhu}
\thanks{shzhu@pku.edu.cn}
\affiliation{Institute of Theoretical Physics $\&$ State Key Laboratory of Nuclear Physics and Technology, Peking University, Beijing 100871, China}
\affiliation{Collaborative Innovation Center of Quantum Matter, Beijing 100871, China}
\affiliation{Center for High Energy Physics, Peking University, Beijing 100871, China}

\date{\today}

\begin{abstract}
In this paper, we discuss the MSSM extended with one vector-like lepton doublets $L$-$\overline{L}$ and one right-handed neutrino $N$. The neutral vecotor-like sneutrino can be a candidate of dark matter. In order to avoid the interaction with the necleons by exchanging a $Z$-boson, the mass splitting between the real part and the imaginary part of the sneutrino field is needed. Compared with the MSSM sneutrino dark matter, the mass splitting between the vector-like sneutrino field can be more naturally acquired without large A-terms and constraints on the neutralino masses. We have also calculated the relic density and the elastic scattering cross sections with the neucleons in the cases that the dark matter particles coannihilate with or without the MSSM slepton doublets. The elastic scattering cross sections with the neucleons are well below the LUX bounds. In the case that the dark matter coannihilate with all the MSSM slepton doublets, the mass of the dark matter can be as light as $370 \text{ GeV}$.

\end{abstract}


\maketitle
\section{Introduction}

In the supersymmetric models, R-parity $(-1)^{(3B+L+2S)}$ usually conserves in order to forbid the protons to decay (For a review, see \cite{Primer}). Then the lightest supersymmetric particle (LSP) can become the dark matter if it is neutral. Neutralinos and sneutrinos have been considered as the candidates of the dark matter in the literature. However, compared with the neutralinos, sneutrinos in the minimal supersymmetric standard model (MSSM) suffer from the difficulty in escaping the direct detection bounds since they can exchange Z-boson with the neucleons \cite{SNeutrinoDMAncestor}. One way to avoid this problem is to introduce the mass-splitting between the real part and the imaginary part of the sneutrino field \cite{Splitting1, Splitting2, Splitting3, Splitting4, Splitting5}. This trick has been applied in many inelastic dark matter models (For examples, see Ref.~\cite{Inelastic1, Inelastic2, Inelastic3, Inelastic4, Inelastic5}). In order to achieve this splitting we need some lepton-number violating sectors beyond the MSSM, which would arise from either the right-handed neutrinos, or some $SU(2)_L$-triplet Higgs fields. These sectors can also make up for the deficiency of the MSSM that the neutrinos are massless. However, in order to acquire the enough splitting value $|m_{\tilde{\nu}^+} - m_{\tilde{\nu}^-}|\gtrsim 100 \text{ KeV}$ and at the same time keep the sub-eV masses of the light neutrinos, large A-terms are usually required, and limits on the masses of the neutralinos are also imposed.

In this paper, we discuss a model that extend the MSSM with a pair of vector-like leptons ($L+\overline{L}$). If the vector-like sneutrinos end up as the dark matter, we also need to split the real part and the imaginary part of the vector-like sneutrino field. The simplest way to achieve this is to introduce another right-handed neutrino field $N$ together with the lepton number violating terms motivated from the type I see-saw mechanisms \cite{SeeSaw1, SeeSaw2, SeeSaw3, SeeSaw4, SeeSaw5}. We will see that in this model, enough mass-splitting can arise from the $L H_u N$ and $\overline{L} H_d N$ Yukawa-terms even if we switch off all the trilinear A-terms. The values of these Yukawa coupling constants can have impact on the relic density of the dark matter, and can also contribute to the direct detection signals. If the mixings between the vector-like sectors and the MSSM sectors are small enough, the sub-eV neutrino masses can also remain undisturbed, relaxing the bounds on the masses of the neutralinos. In the literature, there are models that the MSSM are extended with the vector-like particles (For examples, see Ref.~\cite{Graham:2009gy, Liu:2009cc, Martin:2009bg, Martin:2010dc, Moroi:2011aa, Martin:2012dg, Joglekar:2013zya, Fischler:2013tva, Chang:2013eia, DeSimone:2012fs}). Vector-like sectors can either be heavier than $100 \text{ TeV}$ scale and play the role of so-called ``messengers'' in the gauge mediating supersymmetry breaking (GMSB) models, or influence the TeV-scale phenomenologies if the vector-like particles are relatively light. The later case is particularly interesting partly because TeV-scale vector-like particles can be tested directly through collider searches in the LHC era. Vector-like particles can also interact with the Higgs sectors, relieving the little-hierarchy problem to reach the sufficient standard model (SM)-like Higgs mass in the MSSM. 

We should note that in order to keep the unification of the gauge-coupling constants, our model can be embedded in a $5+\overline{5}$ model, which also contain a pair of vector-like down-type quarks ($D+\overline{D}$). However in the following text, we disregard this. In the Ref.~\cite{NearlyRepeat1, NearlyRepeat2}, there is a similar model that the vector-like messenger sleptons as light as one to three TeV play the role of the dark matter in the framework of the GMSB models (For a review see Ref.~\cite{GMSB_Int}). However, in this paper, we do not concern the origin of the breaking of the supersymmetry, and the vector-like leptons just sense the supersymmetry breaking indirectly, just similar to the ordinary MSSM fields. The dark matter can become much lighter when coannihilating with the MSSM sleptons in our model.

This paper is organized as follows. Section \ref{ModelDescription} describes the model and calculations of the mass matrices are presented. Section \ref{Numerical} calculates the relic density and the spin independent cross section with the neuclons numerically. The Yukawa couplings constants are adjusted in order for a best-fitting to the Planck's result of relic density  \cite{Planck}. Finally, section \ref{Conclusion} contains the conclusions and discussions. 

\section{Model Descriptions} \label{ModelDescription}

Besides the MSSM chiral superfields $H_u$, $H_d$, $L_i$, $E_i$, $Q_i$, $U_i$, $D_i$ ($i=1\text{-}3$), which are the up-type Higgs doublet, down-type Higgs doublet, together with the left-handed lepton doublets, the right-handed charged leptons, the left-handed quark doublets, the up-type and the down-type right-handed quarks of the three generations respectively, we introduce $L$, $\overline{L}$, $N$ in our model, which are a pair of vector-like lepton doublets and one right-handed neutrino. They are assigned with the odd R-parity. The involving superpotential is given by
\begin{eqnarray}
W \supset \mu_L L \overline{L} + y_L L H_u N + y_{\overline{L}} \overline{L} H_d N + \mu_N N^2 + \mu H_u H_d. \label{SuperPotential}
\end{eqnarray}
The supersymmetric breaking soft mass terms and the trilinear A-terms are given by
\begin{eqnarray}
\mathcal{L}_{\text{soft}} &\supset& m_L^2 |\tilde{L}|^2 + m_{\overline{L}}^2 |\tilde{\overline{L}}|^2 + m_N^2 |\tilde{N}|^2 + B_N \mu_N (\tilde{N}^2 + \text{h.c.}) + B_{L} \mu_L ( \tilde{L} \tilde{\overline{L}} + \text{h.c.} ) \nonumber \\
&+& (A_{y_L} y_L \tilde{L} H_u \tilde{N} + A_{y_{\overline{L}}} y_{\overline{L}} \tilde{\overline{L}} H_d \tilde{N} + \text{h.c.} ). \label{SoftTerms}
\end{eqnarray}
Generally speaking, (\ref{SuperPotential}-\ref{SoftTerms}) do not contain all the possible terms which conserve the $U(1)_Y \times SU(2)_L \times SU(3)_C$ quantum numbers and the R-parity. These terms either result in the mixings between the MSSM sectors and the vector-like sectors (e.g., $\tilde{L_i}^{\dagger} \tilde{L}$), or lead to the light-neutrino masses through both the tree-level Type I see-saw mechanisms or loop-level effects  \cite{Loop1, Loop2} (e.g., $y_i L_i H_u N$, together with the corresponding A-terms). In the former case, we assume these terms are small enough to be omitted not only for simplicity, but also because of the precision electroweak constraints on the mixings between the MSSM and the vector-like sectors. For the latter case, the detailed specific mass spectrum and the mixing patterns of the neutrino sectors are out of the scope of this paper, and the smallness of the neutrino masses suppresses the effects from these terms. However, we should note that all these terms cannot be totally absent, because in some coannihilation cases to be discussed, these terms supply the way for the coannihilating particles to finally decay into the dark matter particles.

The conventions of the vacuum expectation values (VEV) of the Higgs sectors are
\begin{eqnarray}
H_u^0 = v_u + \frac{R_u + i I_u}{\sqrt{2}}, ~~~H_d^0 = v_d + \frac{R_d + i I_d}{\sqrt{2}}.
\end{eqnarray}
After the Higgs doublets acquire the VEVs, the real part and the imaginary part of the vector-like neutral sneutrinos are separated. We define
\begin{eqnarray}
\tilde{L} = \left[
\begin{array}{c}
\frac{R_L + i I_L}{\sqrt{2}} \\
\tilde{L}^- 
\end{array} \right], ~~
\tilde{\overline{L}} = \left[
\begin{array}{c}
\tilde{\overline{L}}^+ \\
\frac{R_{\overline{L}} + i I_{\overline{L}}}{\sqrt{2}}
\end{array} \right], ~~
N = \frac{R_N + i I_N}{\sqrt{2}}.
\end{eqnarray}
The mass matrices are therefore
\begin{eqnarray}
V \supset \frac{1}{2}  [R_L,~~R_{\overline{L}},~~R_N] \mathcal{M}_R \left[
\begin{array}{c}
R_L \\
R_{\overline{L}} \\
R_N
\end{array} \right] + \frac{1}{2} [I_L,~~I_{\overline{L}},~~I_N] \mathcal{M}_I \left[
\begin{array}{c}
I_L \\
I_{\overline{L}} \\
I_N
\end{array} \right],
\end{eqnarray}
where
\begin{eqnarray}
\mathcal{M}_R = \mathcal{M}_{RF}+\mathcal{M}_{RD}+\mathcal{M}_{RS}, \nonumber \\
\mathcal{M}_I = \mathcal{M}_{IF}+\mathcal{M}_{ID}+\mathcal{M}_{IS}.
\end{eqnarray}
The matrix elements originating from the F-terms are {\scriptsize
\begin{eqnarray}
\scriptsize \mathcal{M}_{RF} &=& \left[
\begin{array}{ccc}
y_L^2 v_u^2 + \mu_L^2 & -y_L y_{\overline{L}} v_u v_d & - y_L \mu v_d - y_{\overline{L}} v_d \mu_L + 2 y_L v_u \mu_N \\
-y_L y_{\overline{L}} v_u v_d & y_{\overline{L}}^2 v_d^2 + \mu_L^2 & y_{\overline{L}} v_u \mu + y_L \mu_L v_u - 2 y_{\overline{L}} v_d \mu_N \\
-y_L v_d \mu - y_{\overline{L}} v_d \mu_L + 2 y_L \mu_N v_u & y_{\overline{L}} v_u \mu + y_L v_u \mu_L - 2 y_{\overline{L}} v_d \mu_N & y_L^2 v_u^2 + y_{\overline{L}}^2 v_d^2 + 4 \mu_N^2
\end{array} \right], \nonumber \\
\mathcal{M}_{IF} &=& \left[
\begin{array}{ccc}
y_L^2 v_u^2 + \mu_L^2 & -y_L y_{\overline{L}} v_u v_d & y_L v_d \mu - y_{\overline{L}} v_d \mu_L + 2 y_L v_u \mu_N \\
-y_L y_{\overline{L}} v_u v_d & y_{\overline{L}}^2 v_d^2 + \mu_L^2 & -y_{\overline{L}} v_u \mu + y_L v_u \mu_L - 2 y_{\overline{L}} v_d \mu_N \\
y_L v_d \mu - y_{\overline{L}} v_d \mu_L + 2 y_L v_u \mu_N & -y_{\overline{L}} v_u \mu + y_L v_u \mu_L - 2 y_{\overline{L}} v_d \mu_N & y_L^2 v_u^2 + y_{\overline{L}}^2 v_d^2 + 4 \mu_N^2
\end{array} \right]. \label{RIMass_FTerm}
\end{eqnarray} }
The matrix elements induced by the gauge D-terms are
\begin{eqnarray}
\mathcal{M}_{RD,11}&=&\mathcal{M}_{ID,11}=\frac{1}{4}(-g_1^2 v_u^2 + g_1^2 v_d^2 - g_2^2 v_u^2 + g_2^2 v_d^2), \nonumber \\
\mathcal{M}_{RD,22}&=&\mathcal{M}_{ID,22}=\frac{1}{4}(g_1^2 v_u^2 - g_1^2 v_d^2 + g_2^2 v_u^2 - g_2^2 v_d^2), \label{RIMass_DTerm}
\end{eqnarray}
and all the other matrix elements of the $\mathcal{M}_{RD}$ and the $\mathcal{M}_{ID}$ equal 0. $g_{1,2}$ are the $U(1)_Y$ and the $SU(2)_L$ gauge coupling constants respectively. The matrix elements induced by the soft terms are
\begin{eqnarray}
\mathcal{M}_{RS} &=& \left[
\begin{array}{ccc}
m_L^2 & B_L \mu_L & y_L A_{y_L} v_u \\
B_L \mu_L & m_{\overline{L}}^2 & y_{\overline{L}} A_{y_{\overline{L}}} v_d \\
y_L A_{y_L} v_u & y_{\overline{L}} A_{y_{\overline{L}}} v_d & m_N^2 + B_N \mu_N
\end{array} \right], \nonumber \\
\mathcal{M}_{IS} &=& \left[
\begin{array}{ccc}
m_L^2 & -B_L \mu_L & -y_L A_{y_L} v_u \\
-B_L \mu_L & m_{\overline{L}}^2 & -y_{\overline{L}} A_{y_{\overline{L}}} v_d \\
-y_L A_{y_L} v_u & -y_{\overline{L}} A_{y_{\overline{L}}} v_d & m_N^2 - B_N \mu_N
\end{array} \right].
\end{eqnarray}
After diagonalizing $\mathcal{M}_{R,I}$, we acquire three CP-even and CP-odd real scalar particles $R_{1,2,3}$ and $I_{1,2,3}$. They are defined as
\begin{eqnarray}
R_L &=& Z_{R11} R_1 + Z_{R12} R_2 + Z_{R13} R_3, \nonumber \\
R_{\overline{L}} &=& Z_{R21} R_1 + Z_{R22} R_2 + Z_{R23} R_3, \nonumber \\
R_N &=& Z_{R31} R_1 + Z_{R32} R_2 + Z_{R33} R_3, \nonumber \\
I_L &=& Z_{I11} I_1 + Z_{I12} I_2 + Z_{I13} I_3, \nonumber \\
I_{\overline{L}} &=& Z_{I21} I_1 + Z_{22} I_2 + Z_{23} I_3, \nonumber \\
I_N &=& Z_{I31} I_1 + Z_{I32} I_2 + Z_{33} I_3,
\end{eqnarray}
where $Z_{I,Rij}$'s are the matrix elements of the diagonalizing matrices. Without loss of generality, we assign an ascending order of masses among $R_{1,2,3}$ and $I_{1,2,3}$. The mass matrix of the charged vector-like sleptons is
\begin{eqnarray}
V \supset [\tilde{L}^{-*}, \tilde{\overline{L}}^+] \mathcal{M}_{\tilde{L}^{\pm}} \left[
\begin{array}{c}
\tilde{L}^- \\
\tilde{\overline{L}}^{+*}
\end{array} \right],
\end{eqnarray}
where
\begin{eqnarray}
\mathcal{M}_{\tilde{L}^{\pm}} = \mathcal{M}_{\tilde{L}^{\pm}F} + \mathcal{M}_{\tilde{L}^{\pm}D} + \mathcal{M}_{\tilde{L}^{\pm}S}.
\end{eqnarray}
The elements originating from the F-terms are simply
\begin{eqnarray}
\mathcal{M}_{\tilde{L}^{\pm}F11} = \mathcal{M}_{\tilde{L}^{\pm}F11} = \mu_L^2, ~~ \mathcal{M}_{\tilde{L}^{\pm}F12} = \mathcal{M}_{\tilde{L}^{\pm}F21} = 0.
\end{eqnarray}
The elements induced by the D-terms are
\begin{eqnarray}
\mathcal{M}_{\tilde{L}^{\pm}D11} &=& \frac{1}{4} g_1^2 v_d^2 - \frac{1}{4} g_2^2 v_d^2 - \frac{1}{4} g_1^2 v_u^2 + \frac{1}{4} g_2^2 v_u^2 \nonumber \\
\mathcal{M}_{\tilde{L}^{\pm}D22} &=& -\frac{1}{4} g_1^2 v_d^2 + \frac{1}{4} g_2^2 v_d^2 + \frac{1}{4} g_1^2 v_u^2 - \frac{1}{4} g_2^2 v_u^2 \nonumber \\
\mathcal{M}_{\tilde{L}^{\pm}D12} &=& \mathcal{M}_{\tilde{L}^{\pm}D21} = 0. \label{ChargedMass_DTerm}
\end{eqnarray}
The matrix elements induced by the soft terms are
\begin{eqnarray}
\mathcal{M}_{\tilde{L}^{\pm}S11} = \left[
\begin{array}{cc}
m_L^2 & -B_L \mu_L \\
-B_L \mu_L & m_{\overline{L}}^2
\end{array} \right].
\end{eqnarray}
After diagonalizing the $\mathcal{M}_{L^{\pm}}$, we acquire two charged sleptons,
\begin{eqnarray}
\tilde{L}^- = Z_{c11} \tilde{L}_1^- + Z_{c12} \tilde{L}_2^-,~~~\tilde{\overline{L}}^{+*} = Z_{c21} \tilde{L}_1^- + Z_{c22} \tilde{L}_2^-,
\end{eqnarray}
where $Z_{cij}$'s are the diagonalizing matrix elements. The mass matrix of the vector-like neutrinos together with the right-handed neutrino is given by
\begin{eqnarray}
\mathcal{L} \supset \frac{1}{2} [L^{0C}, \overline{L}^{0C}, N^C] \mathcal{M}_{L^0} \left[
\begin{array}{c}
L^0 \\
\overline{L}^{0} \\
N
\end{array} \right],
\end{eqnarray}
where $X^c = X^{\dagger} \cdot (i \sigma^2)$, $\sigma^i$ ($i=1,2,3$) are the Pauli matrices, and $X$ is a two-component Weyl-spinor. The matrix elements of the $\mathcal{M}_{L^0}$ are
\begin{eqnarray}
\mathcal{M}_{L^0} = \left[
\begin{array}{ccc}
0 & \mu_L & y_L v_u \\
\mu_L & 0 & -y_{\overline{L}} v_d \\
y_L v_u & -y_{\overline{L}} v_d & 2 \mu_N
\end{array} \right].
\end{eqnarray}
After diagonalizing the $\mathcal{M}_{L^0}$, we acquire these three neutral majorana fermions,
\begin{eqnarray}
L^0 &=& Z_{011} L^0_1 + Z_{012} L^0_2 + Z_{013} L^0_3, \nonumber \\
\overline{L}^0 &=& Z_{021} L^0_1 + Z_{022} L^0_2 + Z_{023} L^0_3, \nonumber \\
N &=& Z_{031} L^0_1 + Z_{032} L^0_2 + Z_{033} L^0_3,
\end{eqnarray}
where $Z_{0ij}$'s are the diagonalizing matrix elements.

Finally, $L^-$ and $\overline{L}^+$ form a Dirac fermion, and its mass is $\mu_L$.

From observing (\ref{RIMass_FTerm}) we can learn that although $\mathcal{M}_{RF,11} = \mathcal{M}_{IF,11}$, $\mathcal{M}_{RF,22} = \mathcal{M}_{IF,22}$, the off-diagonal $|\mathcal{M}_{RF,13}| \neq |\mathcal{M}_{IF,13}|$. This will split the mass between the $R_i$'s and $I_i$'s even if we switch off all the mass terms induce by the D-terms and the A-terms. In some cases, this difference can be well-estimated. For example, if $m_{\overline{L}}^2, \mu_N^2 \gg m_{L}^2$, the lightest two scalar fields, say $R_1$ and $I_1$, would be dominated by $R_L$ and $I_L$, then
\begin{eqnarray}
m_{R_1}^2 - m_{I_1}^2 &\approx& -\frac{(-y_L \mu v_d - y_{\overline{L}} v_d \mu_L + 2 y_L v_u \mu_N)^2}{4 \mu_N^2} + \frac{(y_L \mu v_d - y_{\overline{L}} v_d \mu_L + 2 y_L v_u \mu_N)^2}{4 \mu_N^2} \nonumber \\
&=& \frac{2 y_L^2 \mu v_d v_u \mu_N - y_L y_{\overline{L}} v_d^2 \mu \mu_L}{\mu_N^2},
\end{eqnarray}
so
\begin{eqnarray}
m_{R_1}-m_{I_1} \approx \frac{1}{\overline{m_{R,I_1}}} \frac{y_L^2 \mu v_d v_u}{\mu_N},
\end{eqnarray}
where $\overline{m_{R,I_1}}$ is the average value of the masses of $R_1$ and $I_1$. For example, if $\mu = 500 \text{ GeV}$, $\tan \beta = \frac{v_u}{v_d} = 15$, $\mu_N = 1 \text{ TeV}$, $y_{\overline{L}} = y_L = 0.1$, and $\overline{m_{R,I_1}} = 400 \text{ GeV}$, then $m_{R_1}-m_{I_1} \approx 20 \text{ MeV}$, which is far beyond the needed $O(100 \text{ KeV})$ in order to escape the direct detection bounds. In this scenario, $I_1$ will be lighter then $R_1$, which means $I_1$ tend to become the dark matter if all the $A_{y_L,y_{\overline{L}}}$, $B_{L, N}$ terms are set zero.

\section{Numerical Results of Relic Abundance and Direct Detection} \label{Numerical}

If $m_L^2 \approx m_{\overline{L}}^2$, the masses of the $I_{L, \overline{L}}$, $R_{L, \overline{L}}$, $L^-$, $\overline{L}^+$ are close to each other and there are large mixings between the neutral and the charged sleptons respectively. In order for a clearer aspect, we assume large difference between the $m_L^2$ and the $m_{\overline{L}}^2$ in this paper to avoid the rather complicated mixings and coannihilating cases. The right-handed (s)neutrino mass terms $m_N^2$, $\mu_N$ are also large enough for the right-handed (s)neutrinos to decouple during the annihilating processes. In this situation, the mixings between the right-handed sneutrinos and the vector-like sneutrinos are also suppressed by their large mass differences. 

According to the (\ref{RIMass_DTerm}, \ref{ChargedMass_DTerm}), the mass terms induced by the D-terms lower the masses of the $R,I_L$ dominated particle and increase the mass of the $L^-$ dominated charged sneutrino, while these terms lower the masses of the $R,I_{\overline{L}}$ dominated particle and give rise to the mass of the $\overline{L}^{+*}$ dominated charged sneutrino. It means that if $m_{\overline{L}}^2 \ll m_L^2$, the masses of the $R,I_{\overline{L}}$ dominated particles tend to be a little heavier than the charged $\overline{L}^{+*}$ dominated particle, leaving us a charged lightest supersymmetric particle (LSP) in most cases. Because of this, we assume $m_L^2 \ll m_{\overline{L}}^2$ in the following text. As has been discussed in the previous section, it means that the LSP will be a CP-odd $I_L$ dominated $I_1$.

A-terms also play roles in the annihilating processes. However, as we have noted, $\tilde{N}$ decouples, so both the effects from the $A_{y_L} y_L \tilde{L} H_u \tilde{N}$ and the $A_{y_{\overline{L}}} y_{\overline{L}} \tilde{\overline{L}} H_d \tilde{N}$ terms are suppressed. Although A-terms also modifies the mass spectrum of the supersymmetric particles, numerical calculations also show that $A_{y_L, y_{\overline{L}}} \sim O(100 \text{ GeV})$ does not influence the final results to a notable extent. According to all these reasons, we set $A_{y_L}=A_{y_{\overline{L}}}=0$ in the following discussions.

For simplicity, we also assume that all the other MSSM sparticles and the exotic Higgs bosons decouple except the Binos ($\tilde{B}$), Winos ($\tilde{W}^{\pm, 0}$) and some $SU(2)_L$ doublet sleptons in some coannihilating cases. We set the masses of all the Binos and Winos to be $m_{\tilde{B}} = m_{\tilde{W}^{\pm, 0}} = 2 \text{ TeV}$. We also set the alignment condition $\beta = \frac{\pi}{2} - \alpha$, where $\alpha$ is the neutral Higgs bosons' mixing angle. This equals to the $m_A \rightarrow \infty$ limit, where $m_A$ is the mass of the CP-odd Higgs boson. We set $\mu_L = 300 \text{ GeV}$ during the calculation, which is safe from the bounds on heavy leptons \cite{PDG}. 

\begin{figure}
\includegraphics[width=3in]{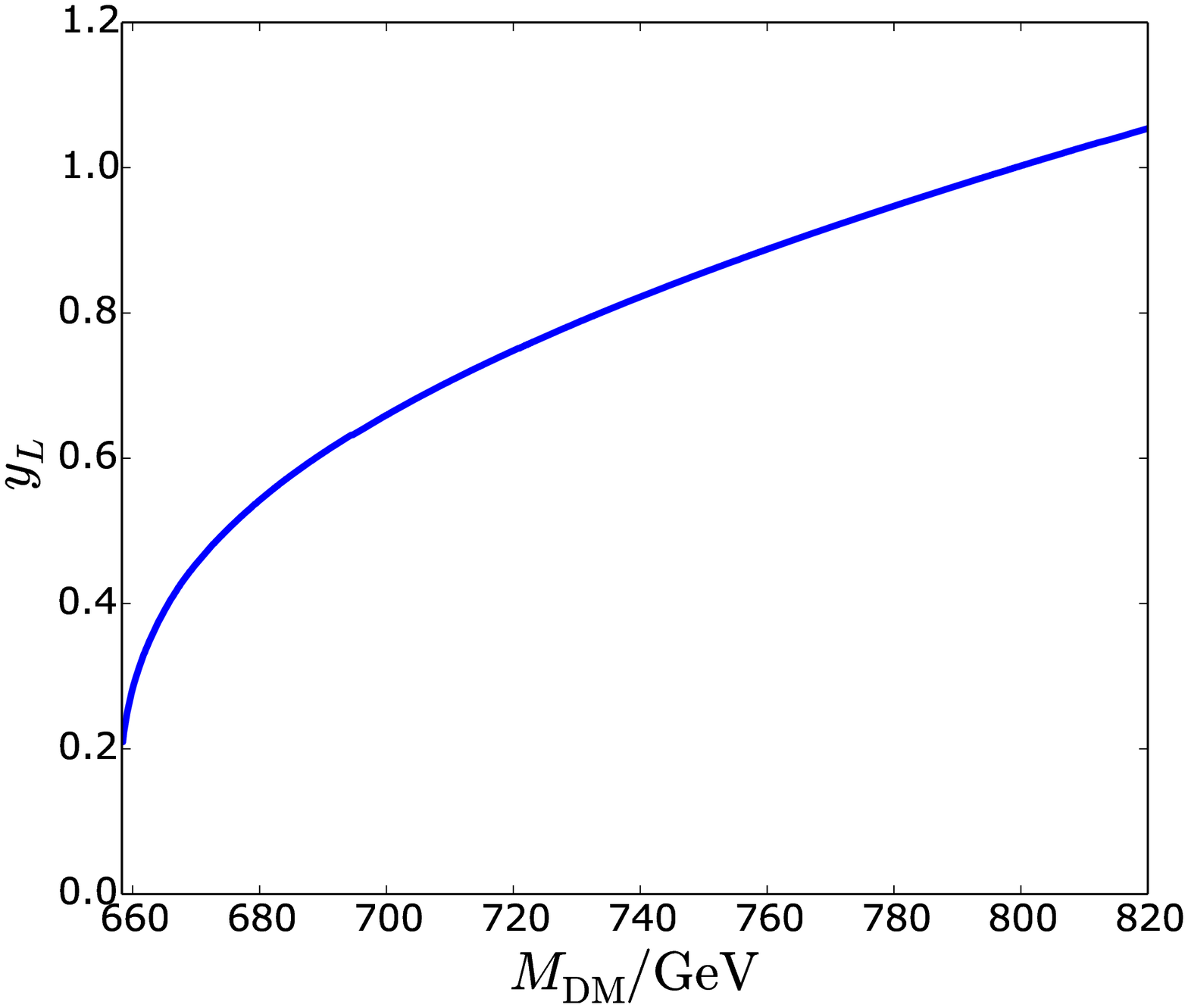}
\includegraphics[width=3in]{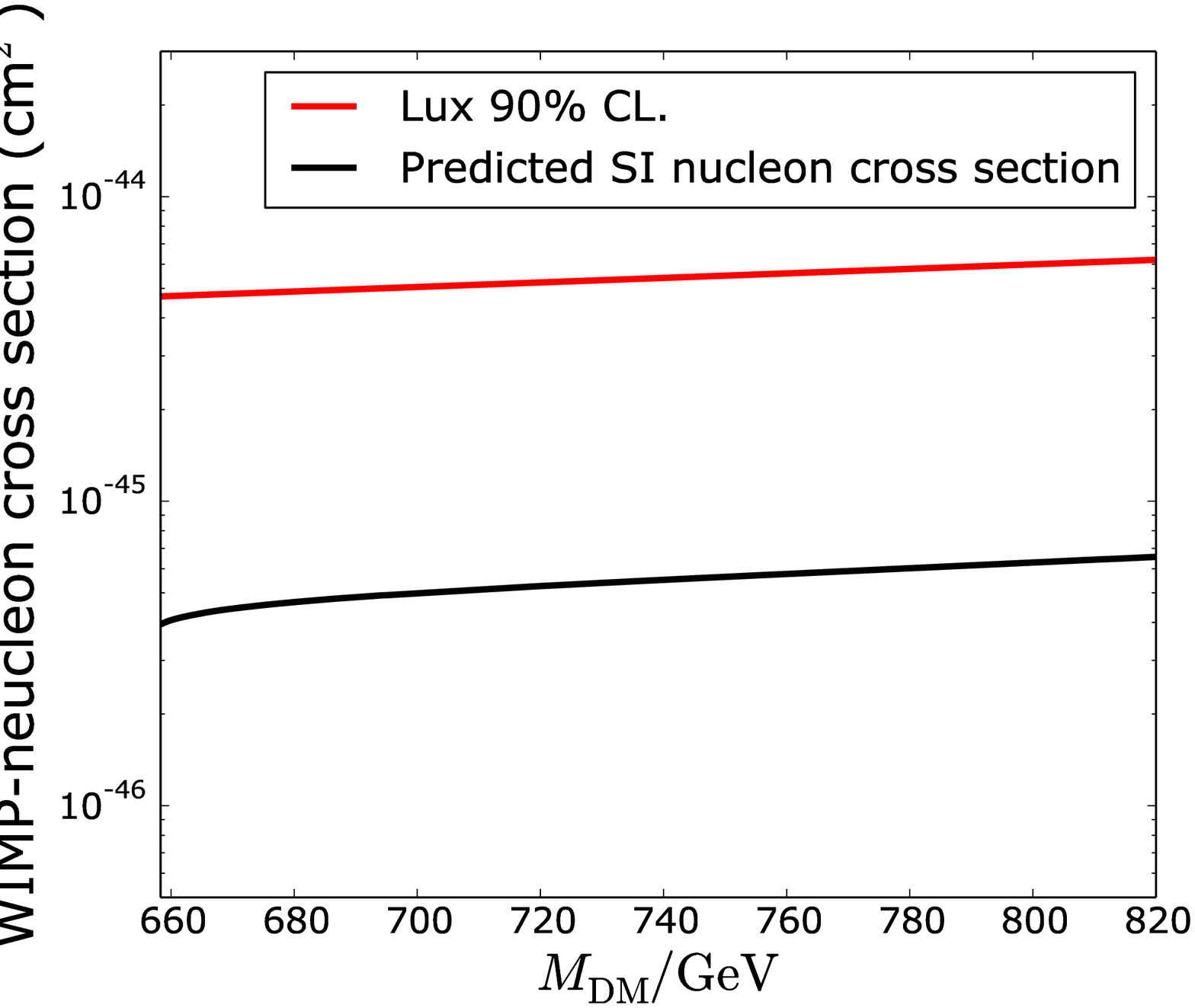}
\includegraphics[width=3in]{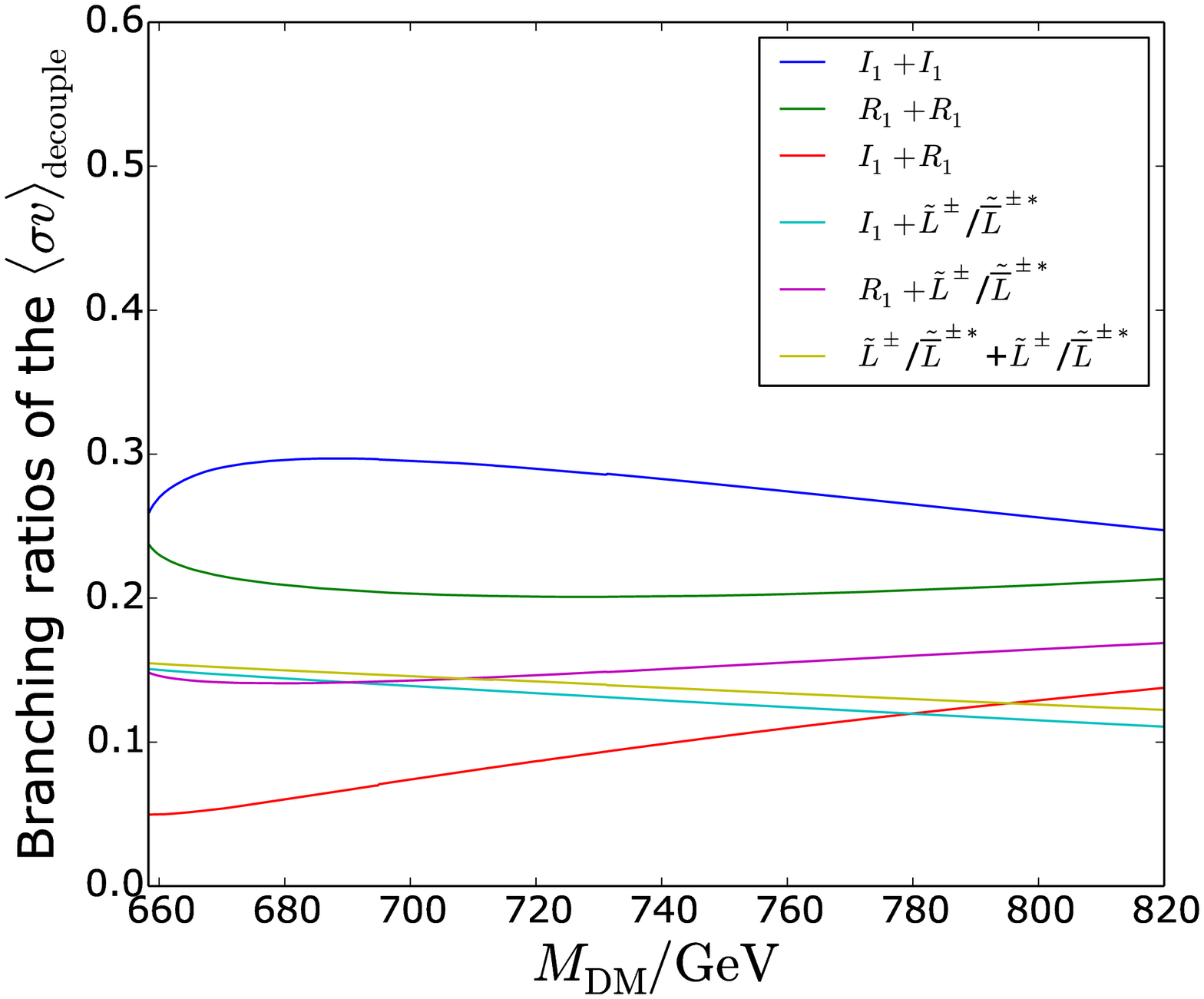}
\caption{The $y_L$ corresponding to $\Omega_c h^2 = 0.1199$ (left pannel), the spin independent cross section with the neucleons of the dark matter particles (right pannel), and the branching ratios of $\langle \sigma v \rangle_{\text{decouple}}$ (bottom pannel) in the case that only the $I_1$, $R_1$, together with $\tilde{L}_1$ coannihilate.}
\label{NoCo}
\end{figure}

\begin{figure}
\includegraphics[width=3in]{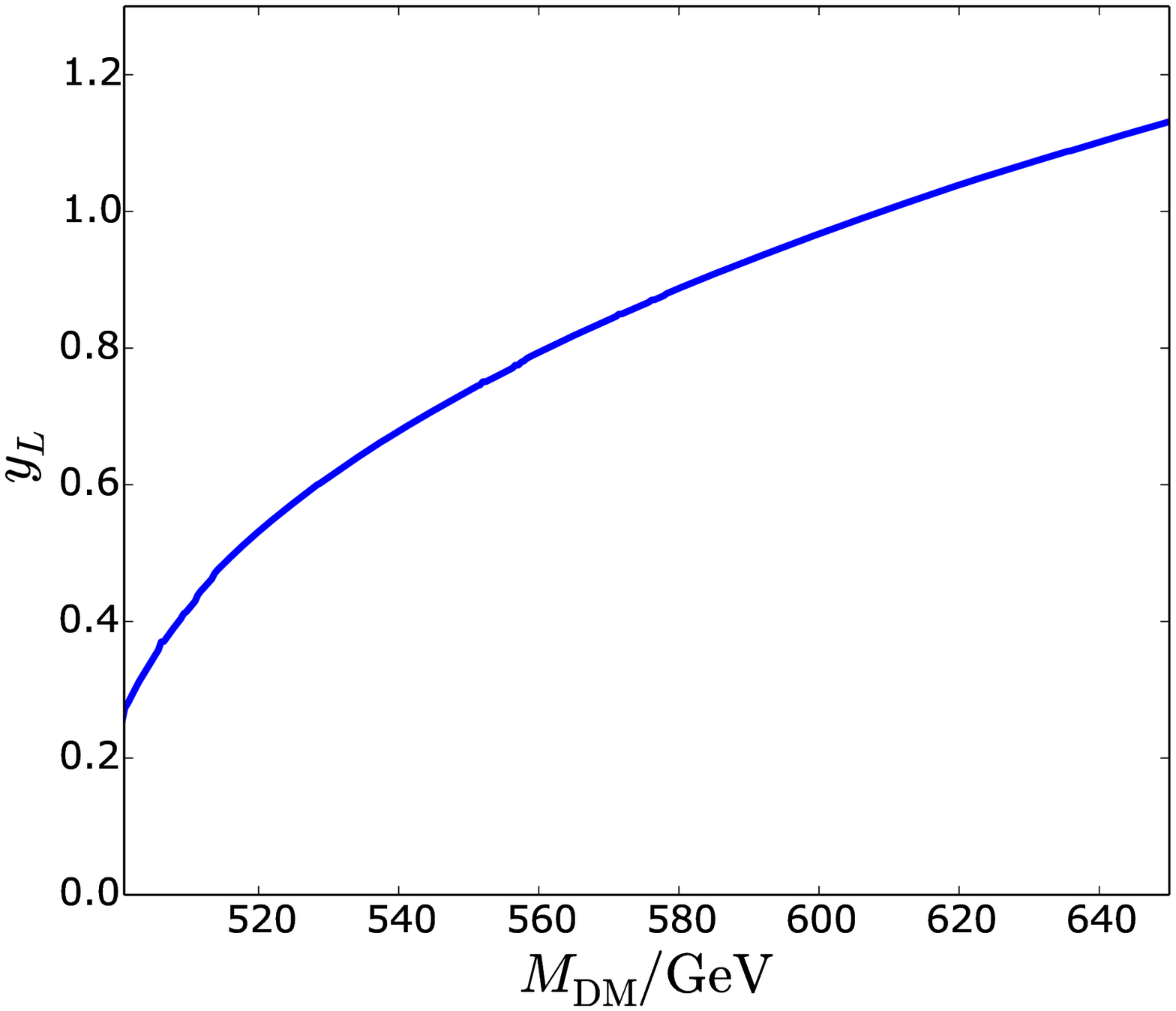}
\includegraphics[width=3in]{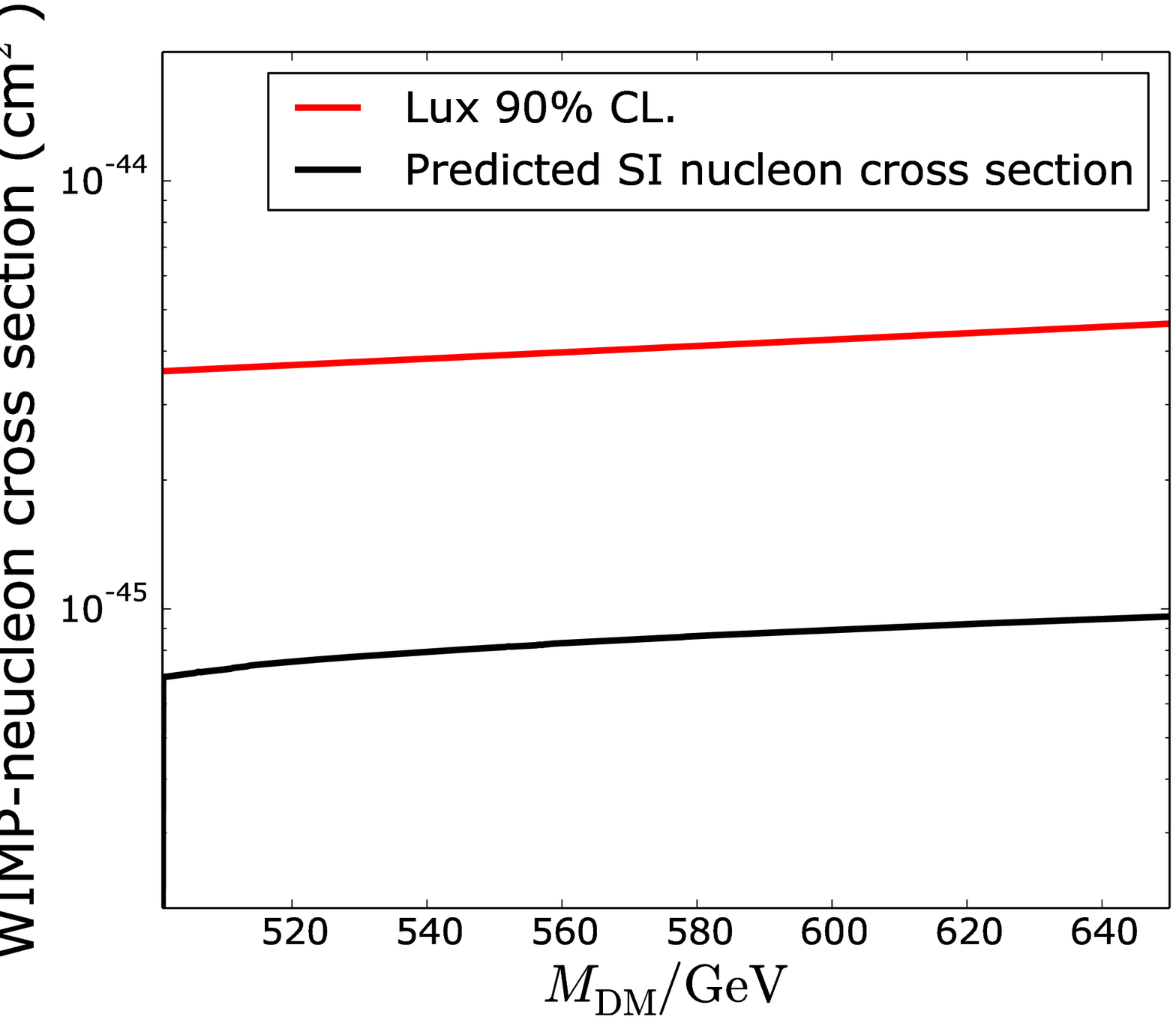}
\includegraphics[width=3in]{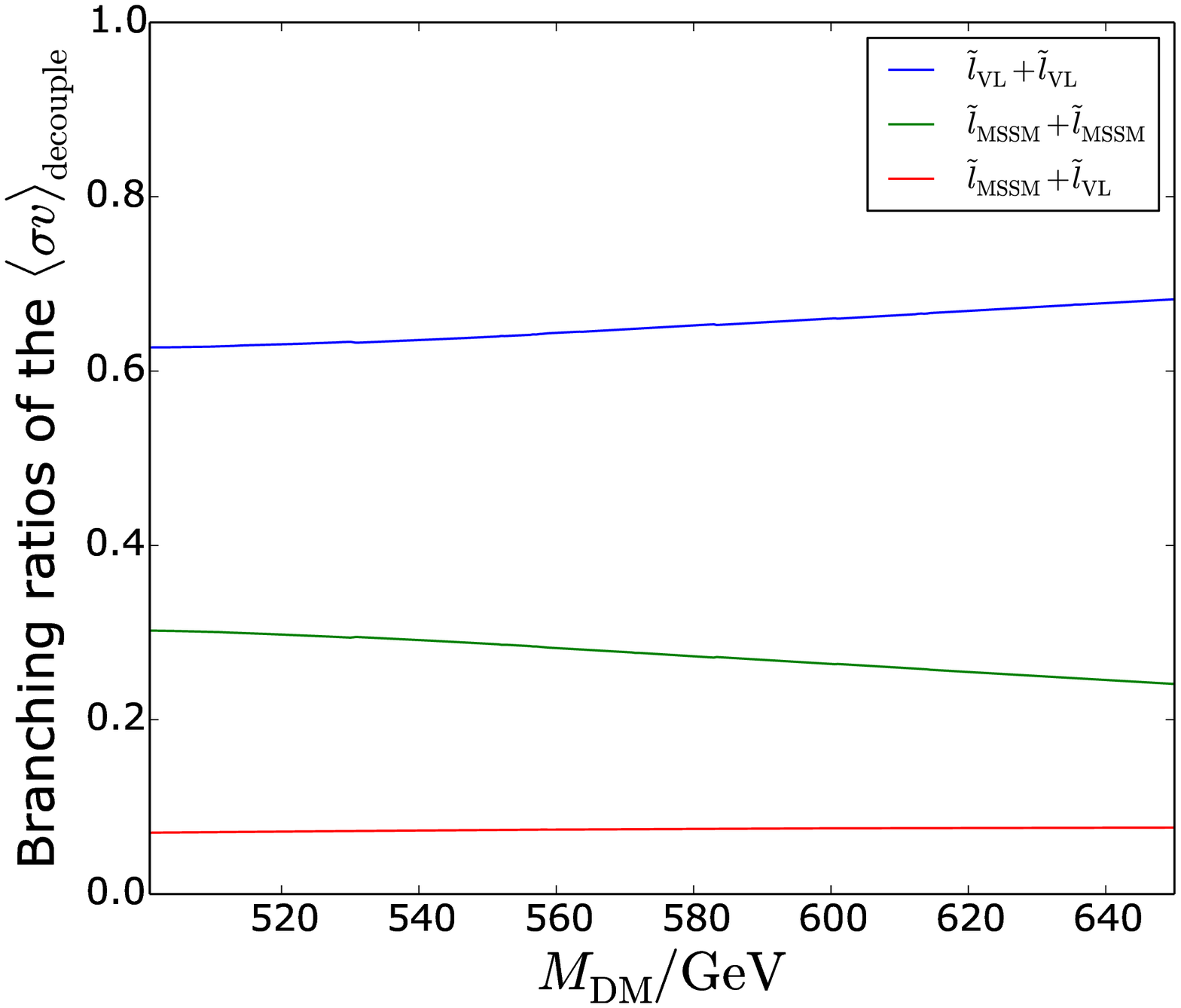}
\caption{The $y_L$ corresponding to $\Omega_c h^2 = 0.1199$ (left pannel), the spin independent cross section with the neucleons of the dark matter particles (right pannel), and the branching ratios of $\langle \sigma v \rangle_{\text{decouple}}$ (bottom pannel) in the case that the vector-like sleptons coannihilate with one generation of MSSM slepton.}
\label{OneCo}
\end{figure}

\begin{figure}
\includegraphics[width=3in]{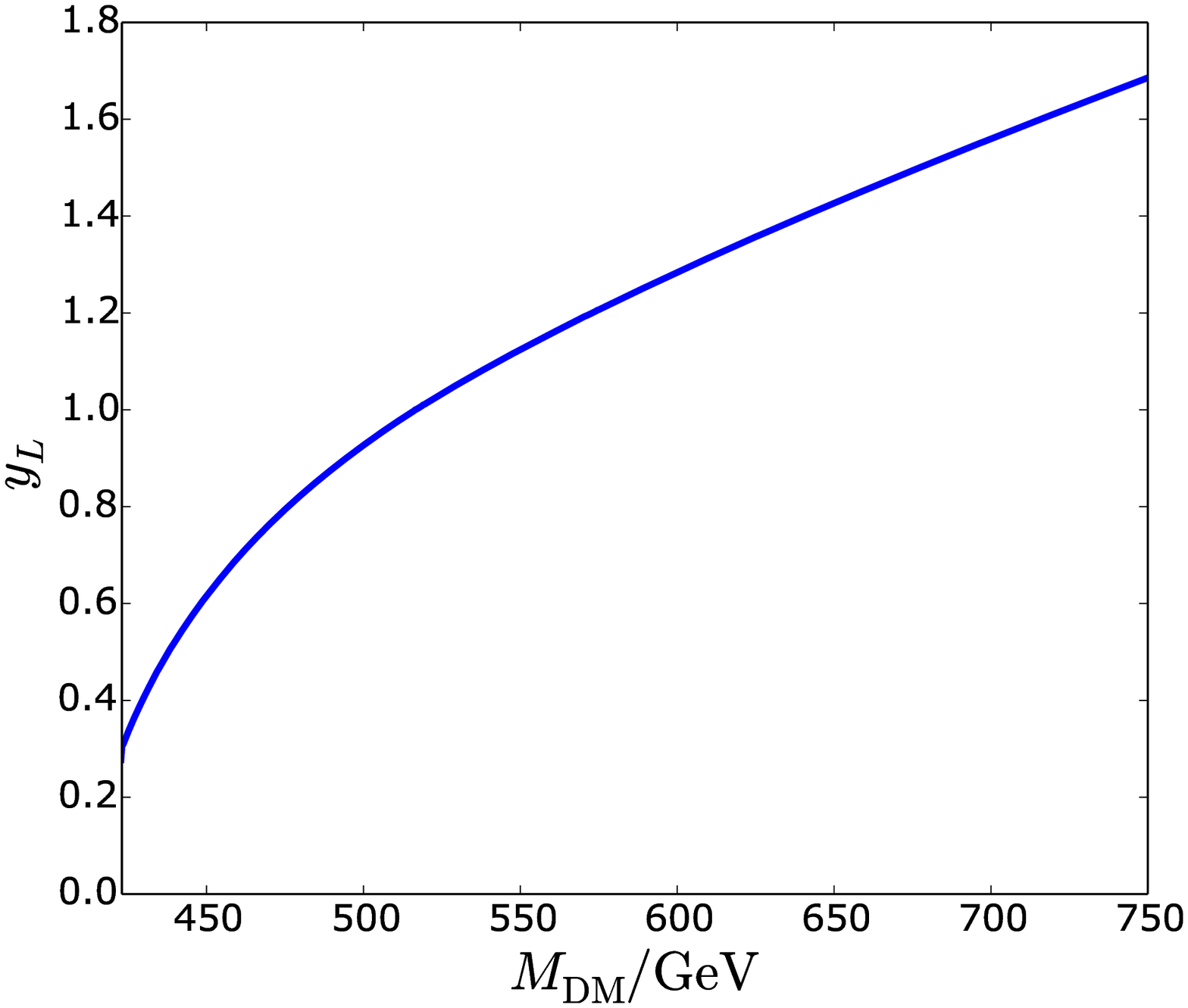}
\includegraphics[width=3in]{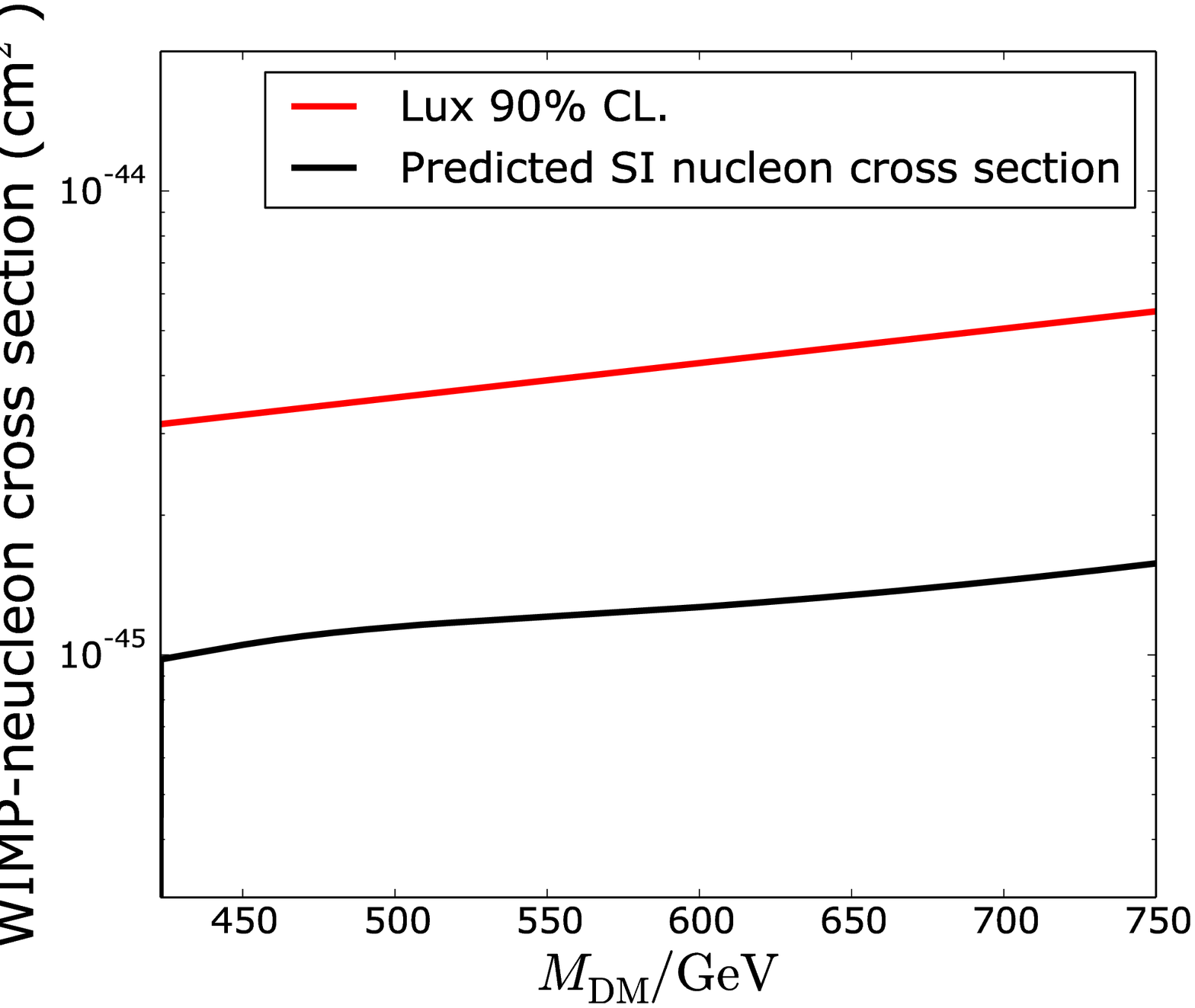}
\includegraphics[width=3in]{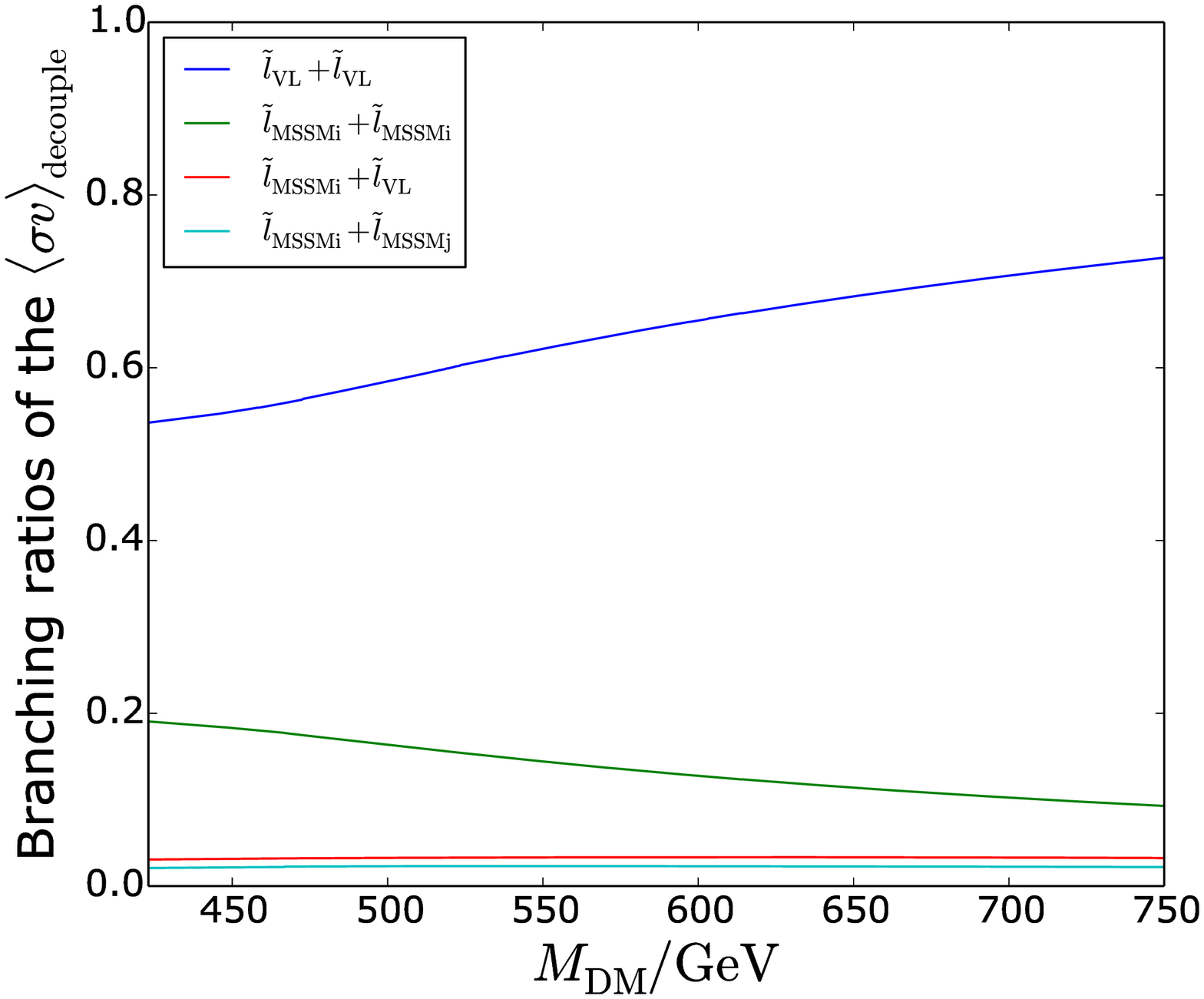}
\caption{The $y_L$ corresponding to $\Omega_c h^2 = 0.1199$ (left pannel), the spin independent cross section with the neucleons of the dark matter particles (right pannel), and the branching ratios of $\langle \sigma v \rangle_{\text{decouple}}$ (bottom pannel) in the case that the vector-like sleptons coannihilate with two generation of MSSM slepton.}
\label{TwoCo}
\end{figure}

\begin{figure}
\includegraphics[width=3in]{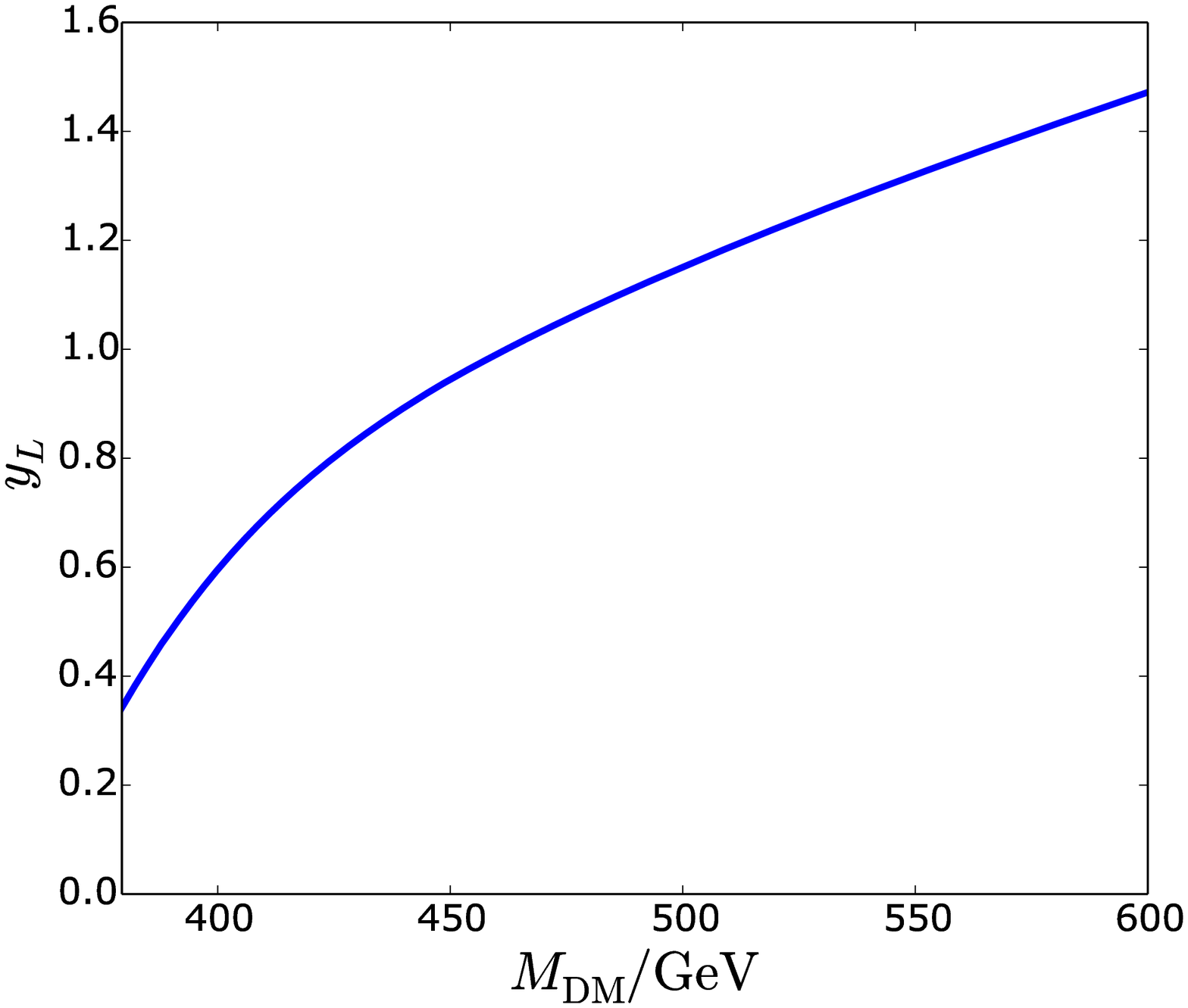}
\includegraphics[width=3in]{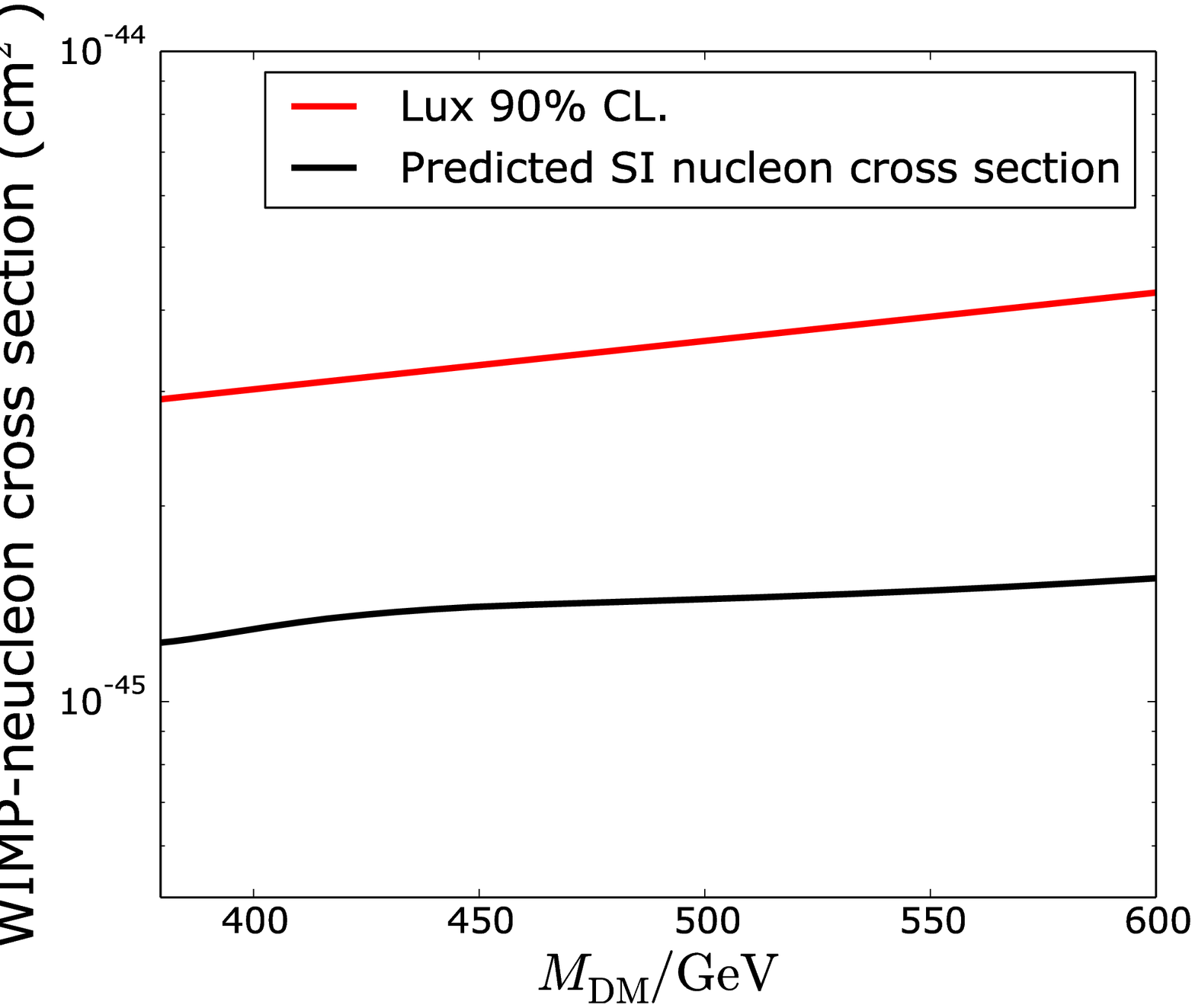}
\includegraphics[width=3in]{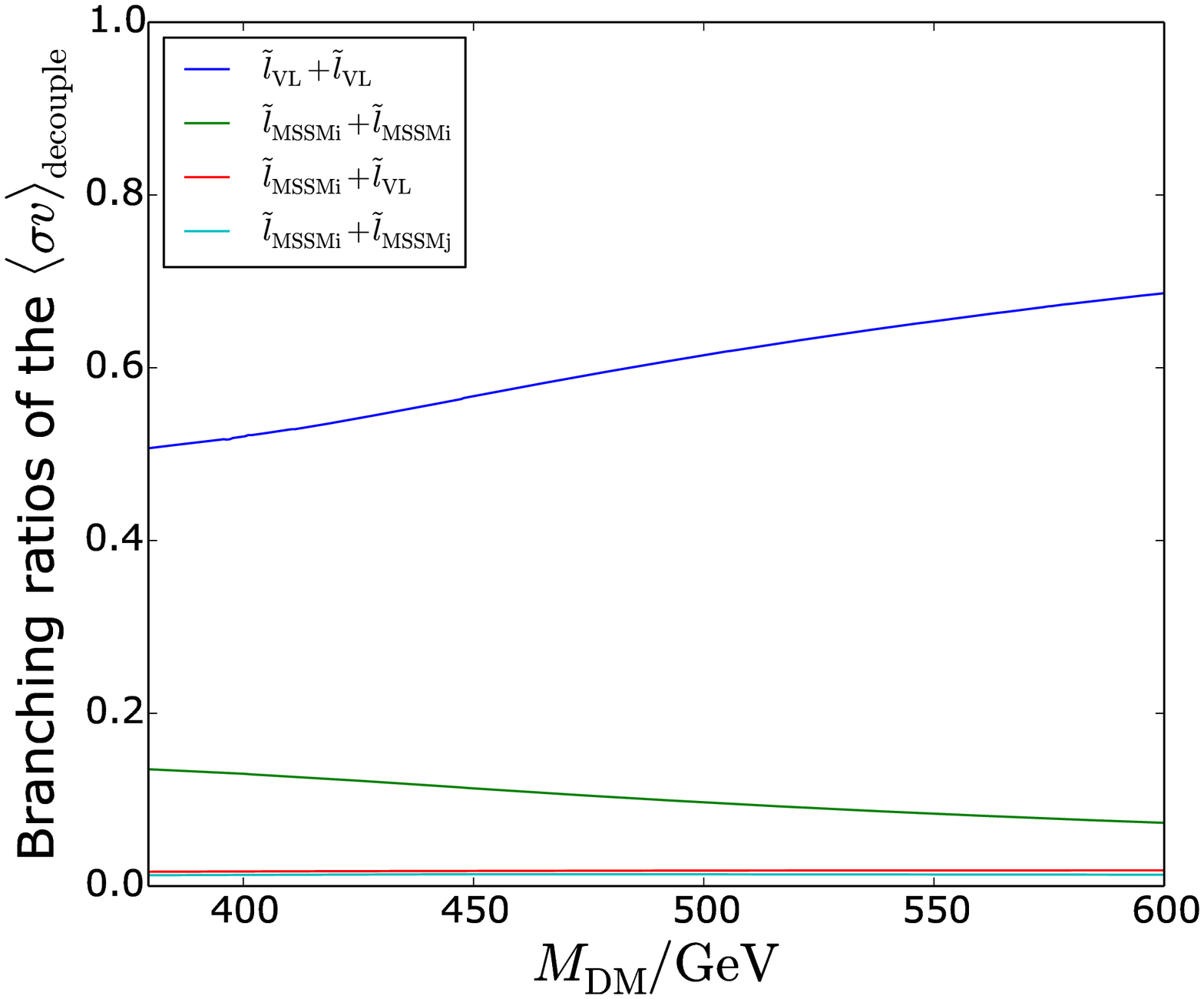}
\caption{The $y_L$ corresponding to $\Omega_c h^2 = 0.1199$ (left pannel), the spin independent cross section with the neucleons of the dark matter particles (right pannel), and the branching ratios of $\langle \sigma v \rangle_{\text{decouple}}$ (bottom pannel) in the case that the vector-like sleptons coannihilate with all the three generation of MSSM slepton.}
\label{FullCo}
\end{figure}

The model is implemented with the FeynRules 2.3.12 \cite{FeynRules} to generate the CalcHEP \cite{CalcHEP} model files. Then MicrOMEGAs 4.2.5 \cite{micrOMEGAs} is used to calculate the relic density, the spin independent cross section with the neucleons, and the branching ratios contributing to the $\langle \sigma v \rangle_{\text{decouple}}$. For each mass of the dark matter, we calculate the $y_L$ which corresponds to the best fitted Planck data $\Omega_c h^2 = 0.1199$ \cite{Planck}, and plot the $m_{\text{DM}}$, $y_L$, branching ratios contributing to the $\langle \sigma v \rangle_{\text{decouple}}$ and the spin independent direct detection cross section with the neucleon $\sigma_{\text{SI}}$ in four cases, which are no coannihilation, coannihilation with one MSSM slepton, coannihilations with two MSSM sleptons, and coannihilations with three MSSM sleptons in the Fig.~\ref{NoCo}, \ref{OneCo}, \ref{TwoCo}, \ref{FullCo}. For each coannihilating situation, we guarantee the masses of the coannihilating MSSM sneutrinos to be 2 GeV heavier than the mass of the dark matter. Note that it is impossible and unnecessary to plot every branching ratio of the $\langle \sigma v \rangle_{\text{decouple}}$ in such small graphs, so we sum over the channels according to the classifications of the initial states. In the Fig.~\ref{NoCo}, we plot the branching ratios among the coannihilating vector-like CP-even/CP-odd sneutrino and the vector-like charged sleptons. In the Fig.~\ref{OneCo}, \ref{TwoCo}, \ref{FullCo}, we only plot the branching ratios among the vector-like sleptons and the MSSM sleptons. If we ignore the masses of the MSSM leptons in our numerical calculations, the branching ratios will become generation-independent, so we only plot one of the branching ratios of each of the $\tilde{l}_{VL} + \tilde{l}_{MSSM i}$, the $\tilde{l}_{MSSM i} + \tilde{l}_{MSSM i}$, and $\tilde{l}_{MSSM i} + \tilde{l}_{MSSM i}(i \neq j)$ in the Fig.~\ref{TwoCo}, \ref{FullCo}.

If the Yukawa coupling constant $y_L$ is switched off, then the main annihilating channels will become the $W^+ W^-$, $ZZ$ channels. The s-channel $R_1 + I_1 \rightarrow Z \rightarrow \overline{l} l$ is suppressed because the $R$-$I$-$Z$ vertex is proportional to $R_1 \partial_{\mu} I_1 - I_1 \partial_{\mu} R_1$. At the decoupling time the four-momentum vector of one dark-side particle is $(m_{\text{DS}} + \frac{1}{2} m_{\text{DS}} v^2, m_{\text{DS}} \vec{v})$. When $v \ll 1$, both terms of $R_1 \partial_{\mu} I_1 - I_1 \partial_{\mu} R_1$ nearly cancel out since $m_{R_1} \approx m_{I_1}$.

Generally speaking, if all the coupling constants stay unchanged, the annihilation cross section $\langle \sigma v \rangle_{\text{decouple}} \propto \frac{1}{m_{\text{DM}}^2}$. If there are only one $I_L$-like $I_1$ together with its companions in the same $SU(2)_L$ doublets, that is to say, the $R_1$, the $\tilde{L}^{-}$ and $\tilde{L}^{+}$ to coannihilate, $m_{\text{DM}} = m_{I_1}$ should be approximately $660$ GeV if $y_L \sim 0$. For a heavier $m_{I_1}$, a larger Yukawa coupling constant $y_L$ is needed in order for a sufficient $\langle \sigma v \rangle_{\text{decouple}} \sim 3 \times 10^{-26} \text{cm}^3/\text{s}$. For a lighter $m_{I_1}$, usually the $\Omega_c h^2$ is suppressed by the too large $\langle \sigma v \rangle_{\text{decouple}}$. This can be improved if the MSSM sleptons coannihilate with the vector-like sleptons. From Fig.~\ref{FullCo} we can see that if the dark matter coannihilate with all the MSSM slepton doublets, $m_{\text{DM}}$ can be as light as $\sim 370 \text{ GeV}$. In the coannihilation scenario, the effective cross section becomes \cite{DarkMatterInt}
\begin{eqnarray}
\langle \sigma_{\text{eff}} v \rangle = \sum_{ij} \langle \sigma_{ij} v_{ij} \rangle \frac{n_i^{\text{eq}}}{n^{\text{eq}}} \frac{n_j^{\text{eq}}}{n^{\text{eq}}},
\end{eqnarray}
where $i$ and $j$ indicate the coannihilating particle content. If $\langle \sigma_{ij} v_{ij} \rangle \ll \langle \sigma_{kk} v_{kk} \rangle$ $(i \neq j)$, then $\langle \sigma_{\text{eff}} v \rangle$ can be suppressed. In this paper, the cross interactions between the vector-like sneutrinos and the MSSM sneutrinos can arise from the exchanges of a t-channel Bino or Wino. Thus, heavier masses of the binos or winos lower the cross interactions and hence lower the $\langle \sigma_{\text{eff}} v \rangle$ effectively in order for the correct relic density in the case of a lighter dark matter. Nevertheless, We should note that the coannihilation scenario requires that $\langle \sigma_{ij} v_{ij} \rangle$ $(i \neq j)$ cannot be too small to avoid the independent annihilation of the different elements, in this case the masses of the Binos and Winos can not be too heavy. As has been mentioned before, we adopt the masses of the Binos and Winos to be $2 \text{ TeV}$, which give rise to the cross interactions plotted in the Fig.~\ref{OneCo}, \ref{TwoCo}, \ref{FullCo}. Further modifying the model can also reach the sufficient cross interactions. For example, in the inverse see-saw model \cite{Inverse1, Inverse2, Inverse3, Inverse4, ZhouYeLing}, the coupling constant $y_i$ in the interaction terms $y_i L_i H_u N$ can be as large as $O(0.1)$, or we can introduce another heavy right-handed neutrino $N^{\prime}$ as heavy as $\sim 10^{12} \text{ GeV}$, then the coupling constants $y_i^{\prime}$, $y_L^{\prime}$ in the interaction terms $y_i^{\prime} L_i H_u N^{\prime}$ and $y_L^{\prime} L H_u N$ can be as large as $O(0.1)$ (For an example, see the discussions in the Appendix B of \cite{HardBreaking}). Both these scenarios result in significant $\tilde{L}^{\dagger} H_u H_u^{\dagger} \tilde{L}_i$ terms to reach sufficient $\langle \sigma_{L^0 L_i^0} v_{L^0 L_i^0}\rangle$ in order to keep them ``co''-annihilating.

As the mass of the dark matter rises up in each coannihilation scenario, the $y_L$ is lifted in order to reach the correct relic density. $y_L$ also contribute to the spin independent cross section of the dark matter with the neucleons. Various experiments \cite{DAMA, LUX, XENON100, XENON10, KIMS1, CDMSII, SUPERCDMS} have been carried out in order to constrain the dark matter parameters. Among them we plot the most stringent bound from the LUX \cite{LUX} in all the Fig.~\ref{NoCo}-\ref{FullCo} in comparison with our predicted data. We can see that although $y_L$ increases as the dark matter mass grows, the constraint line still runs forward the predicted spin independent cross section.

Finally, we are going to point out that in order to avoid the Landau pole before the gauge coupling constants' unification in a complete $5+\overline{5}$ model, $y_L$ should be less then $0.765$. This eliminate much area in Fig.~\ref{NoCo}-\ref{FullCo} when the masses of the dark matter particles are heavy. On the other hand, in this situation the $y_L$ does not make a significant contribution to the SM-like Higgs mass, being unable to relieve the little hierarchy problem. However, if we relax this condition, the corrections to the SM-like Higgs mass is proportional to $y_L^4$. If $y_L \sim 1$, and then $m_{\text{L}} \sim m_{DM}$ is heavy, the Higgs mass can be raised effectively and we can reach a possible solution to the little hierarchy problem. 


\section{Conclusions} \label{Conclusion}

In place of the MSSM sneutrinos, vector-like sneutrinos can play the role of dark matter. Compared with the MSSM sneutrinos, the mass splitting between the real part and the imaginary part of the vector-like sneutrinos can be more naturally acquired without the assumptions of large A-terms and do not bother the light neutrino masses. We have calculated the relic density and the elastic scattering cross section with neucleons of the $I_L$-like dark matter $I_1$. Coannihilating with the MSSM slepton doublets, the dark matter can be as light as $370 \text{ GeV}$. The predicted cross section with neucleons are also below the most stringent experimental bounds from the LUX.

\begin{acknowledgements}

We would like to thank Ran Ding, Jia-Shu Lu, Weihong Zhang, Chen Zhang for helpful discussions.  This work was supported in part by the Natural Science Foundation of China (Grants No.~11135003 and No.~11375014).

\end{acknowledgements}

\newpage
\bibliography{VLSneutrinoDM}
\end{document}